\documentclass[12pt,amssymb,amsmath,tightenlines]{article}
\usepackage{amsmath, amssymb, amsthm, latexsym, mathrsfs}
\usepackage{color}

\newcommand{\N}{\mathbb{N}}

\newcommand{\E}{\mathbb{E}}

\newcommand{\F}{\mathcal{F}}

\theoremstyle{definition}

\numberwithin{equation}{section}
\title{\bf{Discrete-Time Interest Rate Modelling}}
\begin{document}

\author{Lane P. Hughston$^1$ and Andrea Macrina$^{\,2,3}$}
\date{}
\maketitle
\begin{center}
$^1$Department of Mathematics, Imperial College, London SW7 2AZ, UK\\
$^2$Department of Mathematics, King's College London, London, WC2R 2LS, UK\\
$^3$Kyoto Institute of Economic Research, Kyoto University, Kyoto
606-8501, Japan
\end{center}

\begin{abstract}
This paper presents an axiomatic scheme for interest rate models in
discrete time. We take a pricing kernel approach, which builds in
the arbitrage-free property and provides a link to equilibrium
economics. We require that the pricing kernel be consistent with a
pair of axioms, one giving the inter-temporal relations for
dividend-paying assets, and the other ensuring the existence of a
money-market asset. We show that the existence of a positive-return
asset implies the existence of a previsible money-market account. A
general expression for the price process of a limited-liability
asset is derived. This expression includes two terms, one being the
discounted risk-adjusted value of the dividend stream, the other
characterising retained earnings. The vanishing of the latter is
given by a transversality condition. We show (under the assumed
axioms) that, in the case of a limited-liability asset with no
permanently-retained earnings, the price process is given by the
ratio of a pair of potentials.  Explicit examples of discrete-time
models are provided.
\begin{center}
Keywords: Interest rates models; pricing kernels; financial time
series; Flesaker-Hughston models; transversality condition;
financial bubbles.\\
\vspace{.25cm}

Email addresses: lane.hughston@ic.ac.uk, andrea.macrina@kcl.ac.uk\\
Presented at the 7th International ISAAC Congress, London, 13-17
July 2009.
\end{center}
\end{abstract}
%
\section{Discrete-time asset pricing}\label{aba:sec1}
Although discrete-time interest rate models are often introduced for
computational purposes as a convenient approximation to the
continuous-time situation, it is important to recognize that the
theory can be developed in discrete time in an entirely satisfactory
way in its own right, without reference to continuous time. Let
$\{t_i\}_{i=0,1,2,\ldots}$ denote a time sequence where $t_0$ is the
present and $t_{i+1}>t_i$ for all $i\in\N_0$. We assume that
$\{t_i\}$ is unbounded in the sense that for any $T$ there exists a
value of $i$ such that $t_i>T$. The economy is represented by a
probability space $(\Omega, \mathcal{F}, \mathbb{P})$ with a
filtration $\{\mathcal{F}_i\}_{i\ge 0}$ which we call the ``market
filtration". Each asset is characterised by a pair
$\{S_{t_i}\}_{i\ge 0}$ and $\{D_{t_i}\}_{i\ge 0}$ which we call the
``value process" and the ``dividend process". We interpret $D_{t_i}$
as the random cash flow paid out by the asset at time $t_i$. Then
$S_{t_i}$ denotes the ``ex-dividend" value of the asset at $t_i$.
For simplicity, we often write $S_i=S_{t_i}$ and $D_i=D_{t_i}$. To
ensure the absence of arbitrage, we assume the existence of a
positive pricing kernel $\{\pi_i\}_{i\ge 0}$, and make the following
assumptions:

{\bf Axiom~A}. {\it For any asset with value process
$\{S_i\}_{i\ge\infty}$ and dividend process $\{D_i\}_{i\ge 0}$, the
process $\{M_i\}_{i\ge 0}$ defined by
\begin{equation}\label{2.1}
M_i=\pi_i S_i+\sum^{i}_{n=0}\pi_n D_n
\end{equation}
is a martingale.}

{\bf Axiom~B}. {\it There exists a positive non-dividend-paying
asset with value process $\{\bar{B}_i\}_{i\ge 0}$ such that
$\bar{B}_{i+1}>\bar{B}_i$ for all $i\in\mathbb{N}_0$, and that for
any $b\in\mathbb{R}$ there exists a time $t_i$ such that
$\bar{B}_i>b$.}

The notation $\{\bar{B}_i\}$ distinguishes the positive return asset
from the previsible money-market account $\{B_i\}$ introduced later.
Since $\{\bar{B}_i\}$ is non-dividend paying, Axiom A implies that
$\{\pi_i \bar{B}_i\}$ is a martingale. Writing $\bar{\rho}_i=\pi_i
\bar{B}_i$, we have $\pi_i=\bar{\rho}_i/\bar{B}_i$. Since
$\{\bar{B}_i\}$ is increasing, $\{\pi_i\}$ is a supermartingale, and
it follows from Axiom B that
\begin{equation}\label{3.5aa}
\lim_{i\rightarrow\infty}\E[\pi_i]=0.
\end{equation}
We obtain the following result concerning limited-liability assets.

{\bf Proposition 1}. {\it Let $S_i\ge0$ and $D_i\ge0$ for all
$i\in\mathbb{N}$. We have
\begin{equation}\label{2.2}
S_i=\frac{m_i}{\pi_i}+\frac{1}{\pi_i}\E_i\left[\sum^{\infty}_{n=i+1}\pi_n
D_n\right],
\end{equation}
where $\{m_i\}$ is a non-negative martingale that vanishes if and
only if:
\begin{equation}\label{2.3}
\lim_{j\rightarrow\infty}\E[\pi_j S_j]=0.
\end{equation}}
\indent Thus $\{m_i\}$ represents that part of the value of the
asset that is ``never paid out''. An idealised money-market account
is of this nature, and so is a ``permanent bubble'' (cf. Tirole
1982, Armerin 2004). In the case of an asset for which the
``transversality'' condition (\ref{2.3}) is satisfied, the price is
directly related to the future dividend flow:
\begin{equation}\label{2.12}
S_i=\frac{1}{\pi_i}\E_i\left[\sum^{\infty}_{n=i+1}\pi_n D_n\right].
\end{equation}
This is the so-called ``fundamental equation" often used as a basis
for asset pricing (cf. Cochrane 2005). Alternatively we can write
\begin{equation}\label{2.13}
S_i=\frac{1}{\pi_i}(\E_i[F_{\infty}]-F_i),
\end{equation}
where $F_i=\sum^i_{n=0}\pi_n D_n$,and
$F_{\infty}=\lim_{i\rightarrow\infty}F_i$. Hence the price of a pure
dividend-paying asset can be expressed as a ratio of potentials,
giving us a discrete-time analogue of a result of Rogers 1997.
\section{Positive-return asset and pricing kernel}\label{aba:sec2}
Let us introduce the notation
\begin{equation}\label{3.1}
\bar{r}_i=\frac{\bar{B}_i-\bar{B}_{i-1}}{\bar{B}_{i-1}}
\end{equation}
for the rate of return of the positive-return asset realised at time
$t_i$ on an investment made at $t_{i-1}$. The notation $\bar{r}_i$
is used to distinguish the rate of return on $\{\bar{B}_i\}$ from
the rate of return $r_i$ on the money market account $\{B_i\}$
introduced later.

{\bf Proposition 2}. {\it There exists an asset with constant value
$S_i=1$ for all $i\in\N_0$, for which the associated cash flows are
given by $\{\bar{r}_i\}_{i\ge 1}$.}

{\bf Proposition 3}. {\it Let $\{\bar{B}_i\}$ be a positive-return
asset satisfying Axioms A and B, and let $\{\bar{r}_i\}$ be its
rate-of-return process. Then the pricing kernel can be expressed in
the form $\pi_i=\E_i[G_{\infty}]-G_i$, where
$G_i=\sum^i_{n=1}\pi_n\bar{r}_n$ and
$G_{\infty}=\lim_{i\rightarrow\infty}G_i$.}

There is a converse to this result that allows one to construct a
system satisfying Axioms A and B from a strictly-increasing
non-negative adapted process that converges and satisfies an
integrability condition:

{\bf Proposition 4}. {\it Let $\{G_i\}_{i\ge 0}$ be a
strictly-increasing adapted process with $G_0=0$, and
$\E[G_{\infty}]<\infty$, where
$G_{\infty}=\lim_{i\rightarrow\infty}G_i$. Let $\{\pi_i\}$,
$\{\bar{r}_i\}$, and $\{\bar{B}_i\}$ be defined by
$\pi_i=\E_i[G_{\infty}]-G_{i}$ for $i\ge 0$,
$\bar{r}_i=(G_i-G_{i-1})/\pi_i$ for $i\ge 1$, and
$\bar{B}_i=\prod^i_{n=1}(1+\bar{r}_n)$ for $i\ge 1$, with
$\bar{B}_0=1$. Let $\{\bar{\rho}_i\}$ be defined by
$\bar{\rho}_i=\pi_i \bar{B}_i$ for $i\ge0$. Then $\{\bar{\rho}_i\}$
is a martingale, and $\lim_{j\rightarrow\infty}\bar{B}_j=\infty$,
from which it follows that $\{\pi_i\}$ and $\{\bar{B}_i\}$ satisfy
Axioms A and B.}
\section{Discrete-time discount bond systems}\label{aba:sec3}
The price $P_{ij}$ at $t_i$ of a discount bond that matures at $t_j$
$(i<j)$ is
\begin{equation}
P_{ij}=\frac{1}{\pi_i}\E_i[\pi_j].
\end{equation}
Since $\pi_i>0$ for $i\in\N$, and $\E_i[\pi_j]<\pi_i$ for $i<j$, it
follows that $0<P_{ij}<1$ for $i<j$. We observe that the
``per-period'' interest rate $R_{ij}$ defined by
$P_{ij}=1/(1+R_{ij})$ is positive. Since $\{\pi_i\}$ is given, there
is no need to model the volatility structure of the bonds. Thus, our
scheme differs from the discrete-time models discussed in Heath {\it
et al.} 1990, and Filipovi\'c \& Zabczyk 2002. As an example of a
class of discrete-time models set $\pi_i=\alpha_i+\beta_i N_i$,
where $\{\alpha_i\}$ and $\{\beta_i\}$ are positive,
strictly-decreasing deterministic processes with
$\lim_{i\rightarrow\infty}\alpha_i=0$ and
$\lim_{i\rightarrow\infty}\beta_i=0$, and where $\{N_i\}$ is a
positive martingale. Then we have
\begin{eqnarray}\label{3.18}
P_{ij}=\frac{\alpha_j+\beta_j N_i}{\alpha_i+\beta_i N_i},
\end{eqnarray}
giving a family of ``rational" interest rate models. In a
discrete-time setting we can produce models that do not necessarily
have analogues in continuous time---for example, we can let
$\{N_i\}$ be the martingale associated with a branching process.

Any discount bond system consistent with our scheme admits a
representation of the Flesaker-Hughston type (Rutkowski 1997, Jin \&
Glasserman 2001, Cairns 2004, Musiela \& Rutkowski 2005, Bj\"ork
2009). More precisely, we have the following:

{\bf Proposition 5}. {\it Let $\{\pi_i\}$, $\{\bar{B}_i\}$,
$\{P_{ij}\}$ satisfy {\rm Axioms A} and {\rm B}. Then there exists a
family of positive martingales $\{m_{in}\}_{0\le i\le n}$, $n\in\N$,
such that
\begin{equation}\label{3.19}
P_{ij}=\frac{\sum^{\infty}_{n=j+1}m_{in}}
{\sum^{\infty}_{n=i+1}m_{in}}.
\end{equation}}
\section{Construction of the money-market asset}\label{aba:sec4}
Let us look now at the situation where the positive-return asset is
previsible. Thus we assume that $B_i$ is $\F_{i-1}$-measurable and
we drop the ``bar" over $B_i$. In that case we have
\begin{eqnarray}\label{3.23}
P_{i-1,i}=\frac{1}{\pi_{i-1}}\E_{i-1}[\pi_i]
=\frac{B_{i-1}}{\rho_{i-1}}\E_{i-1}\left[\frac{\rho_i}{B_i}\right]=\frac{B_{i-1}}{B_i}.
\end{eqnarray}
Hence, writing $P_{i-1,i}=1/(1+r_i)$ where $r_i=R_{i-1,i}$ we see
that the rate of return on the money-market account is previsible,
and is given by the one-period discount factor associated with the
bond that matures at $t_i$.

Reverting to the general situation, it follows that if we are given
a pricing kernel $\{\pi_i\}$ on $(\Omega,\F,\mathbb{P},\{\F_i\})$,
and a system of assets satisfying Axioms A and B, then we can
construct a {\it candidate} for a previsible money market account by
setting $B_0=1$ and
\begin{equation}\label{3.25}
B_i=(1+r_i)(1+r_{i-1})\cdots(1+r_1),
\end{equation}
for $i\ge 1$, where $r_i$ is defined by
\begin{equation}
r_i=\frac{\pi_{i-1}}{\E_{i-1}[\pi_i]}-1.
\end{equation}
We refer to $\{B_i\}$ as the ``natural" money-market account
associated with $\{\pi_i\}$. To justify this terminology, we verify
that $\{B_i\}$, so constructed, satisfies Axioms A and B. To this
end, we note the following multiplicative decomposition. Let
$\{\pi_i\}$ be a positive supermartingale satisfying
$\E_i[\pi_j]<\pi_i$ for $i<j$ and $\lim_{j\rightarrow\infty}
[\pi_j]=0$. Then we can write $\pi_i=\rho_i/B_i$, where
\begin{equation}\label{3.27}
\rho_i=\frac{\pi_i}{\E_{i-1}[\pi_i]}\
\frac{\pi_{i-1}}{\E_{i-2}[\pi_{i-1]}}\cdots\frac{\pi_1}{\E_0[\pi_1]}\pi_0
\end{equation}
for $i\ge 0$, and
\begin{equation}\label{3.28}
B_i=\frac{\pi_{i-1}}{\E_{i-1}[\pi_i]}\
\frac{\pi_{i-2}}{\E_{i-2}[\pi_{i-1}]}\cdots\frac{\pi_1}{\E_1[\pi_2]}\
\frac{\pi_0}{\E_0[\pi_1]}
\end{equation}
for $i\ge 1$, with $B_0=1$. In this scheme we have
\begin{equation}\label{3.29}
\rho_i=\frac{\pi_i}{\E_{i-1}[\pi_i]}\rho_{i-1},
\end{equation}
with $\rho_0=\pi_0$; and
\begin{equation}\label{3.30}
B_i=\frac{\pi_{i-1}}{\E_{i-1}[\pi_i]}B_{i-1},
\end{equation}
with $B_0=1$. It is evident that $\{\rho_i\}$ is $\{\F_i\}$-adapted,
and that $\{B_i\}$ is previsible and increasing. We establish the
following:

{\bf Proposition 6}. {\it Let $\{\pi_i\}$ be a non-negative
supermartingale such that $\E_i[\pi_j]<\pi_i$ for all $i<j\in\N_0$,
and $\lim_{i\rightarrow\infty}\E[\pi_i]=0$. Let $\{B_i\}$ be defined
by $B_0=1$ and $B_i=\prod^i_{n=1}(1+r_n)$ for $i\ge 1$, where
$1+r_i=\pi_{i-1}/\E_{i-1}[\pi_i]$, and set $\rho_i=\pi_i B_i$ for
$i\ge 0$. Then $\{\rho_i\}$ is a martingale, and the interest rate
system defined by $\{\pi_i\}$, $\{B_i\}$, $\{P_{ij}\}$ satisfies
{\rm Axioms A} and {\rm B}. }

A significant feature of Proposition 6 is that no integrability
condition is required on $\{\rho_i\}$: the natural money market
account defined above ``automatically" satisfies Axiom A. Thus in
place of Axiom B we can assume:

{\bf Axiom~B$^{\ast}$}. {\it There exists a positive non-dividend
paying asset, the money-market account $\{B_i\}_{i\ge 0}$, having
the properties that $B_{i+1}>B_i$ for $i\in\N_0$, that $B_i$ is
$\F_{i-1}$-measurable for $i\in\N$, and that for any
$b\in\mathbb{R}$ there exists a $t_i$ such that $B_i>b$.}

The content of Proposition 6 is that Axioms A and B together are
equivalent to Axioms A and B$^{\ast}$ together. Let us establish
that the class of interest rate models satisfying Axioms ${\text A}$
and ${\text B^{\ast}}$ is non-vacuous. In particular, consider the
``rational" model defined for some choice of $\{N_i\}$. It is an
exercise to see that the previsible money market account is given
for $i=0$ by $B_0=1$ and for $i\ge 1$ by
\begin{equation}
B_i=\prod^i_{n=1}\frac{\alpha_{n-1}+\beta_{n-1}
N_{n-1}}{\alpha_n+\beta_n N_{n-1}},
\end{equation}
and that for $\{\rho_i\}$ we have
\begin{equation}
\rho_i=\rho_0\prod^{i}_{n=1}\frac{\alpha_n+\beta_n
N_n}{\alpha_n+\beta_n N_{n-1}},
\end{equation}
where $\rho_0=\alpha_0+\beta_0 N_0$. One can check for each $i\ge 0$
that $\rho_i$ is bounded; therefore $\{\rho_i\}$ is a martingale,
and $\{B_i\}$ satisfies Axioms A and B$^{\ast}$.
\section{Doob decomposition}\label{aba:sec5}

Consider now the Doob decomposition given by
$\pi_i=\E_i[A_{\infty}]-A_i$, with
\begin{eqnarray}
A_i&=&\sum^{i-1}_{n=0}\left(\pi_n-\E_n[\pi_{n+1}]\right)
\end{eqnarray}
as discussed, e.g., in Meyer 1966. It follows that
\begin{equation}
A_i=\sum^{i-1}_{n=0}\pi_n\left(1-\frac{\E_n[\pi_{n+1}]}{\pi_n}\right)=\sum^{i-1}_{n=0}\pi_n\left(1-P_{n,n+1}\right)=\sum^{i-1}_{n=0}\pi_n
r_{n+1}P_{n,n+1},
\end{equation}
where $\{r_i\}$ is the previsible short rate process. The pricing
kernel can therefore be put in the form
\begin{equation}\label{modPK}
\pi_i=\E_i\left[\sum^{\infty}_{n=i}\pi_n r_{n+1}P_{n,n+1}\right].
\end{equation}
Comparing (\ref{modPK}) with the decomposition
$\pi_i=\E_i[G_{\infty}]-G_i$, $G_i=\sum^i_{n=1}\pi_n\bar{r}_n$,
given in Proposition 3, we see that by setting
\begin{equation}
\bar{r}_i=\frac{r_i\pi_{i-1}P_{i-1,i}}{\pi_i}
\end{equation}
we obtain a positive-return asset based on the Doob decomposition.
\section{Foreign exchange processes}
An extension of the material presented here to models for foreign
exchange and inflation is pursued in Hughston \& Macrina (2008). In
particular, since the money-market account is a positive-return
asset, by Proposition 3 we can write:
\begin{equation}
\pi_i=\E_i\left[\sum^{\infty}_{n=i+1}\pi_n r_n\right].
\end{equation}
As a consequence, we see that the price process of a pure
dividend-paying asset can be written in the following symmetrical
form:
\begin{equation}
S_i=\frac{\E_i\left[\sum^{\infty}_{n=i+1}\pi_n
D_n\right]}{\E_i\left[\sum^{\infty}_{n=i+1}\pi_n r_n\right]}.
\end{equation}
In the case where $\{S_i\}$ represents a foreign currency, the
dividend process is the foreign interest rate, and both $\{D_i\}$
and $\{r_i\}$ are previsible.

\end{document}